\documentclass[conference]{IEEEtran}
\IEEEoverridecommandlockouts
\usepackage[comma,sort&compress,numbers]{natbib}
\usepackage{amsmath,amssymb,amsfonts}
\usepackage{algorithmic}
\usepackage{graphicx}
\usepackage{textcomp}
\usepackage{multirow}
\usepackage{dsfont}
\usepackage{float}
\usepackage{booktabs} 
\usepackage{multirow}
\usepackage{siunitx}
\usepackage{hyperref}
\usepackage{url}
\usepackage[table]{xcolor}

\definecolor{beige-color}{RGB}{235, 231, 199} %

\def\BibTeX{{\rm B\kern-.05em{\sc i\kern-.025em b}\kern-.08em
    T\kern-.1667em\lower.7ex\hbox{E}\kern-.125emX}}
    
\begin{document}

\title{Accelerating Learnt Video Codecs with Gradient Decay and Layer-wise Distillation\thanks{\noindent\IEEEauthorrefmark{1}Equal contribution. 
\\
\indent This work is funded by the University of Bristol and the UKRI MyWorld Strength in Places Programme (SIPF00006/1).}}

\author{\IEEEauthorblockN{Tianhao Peng\textsuperscript{\dag}\IEEEauthorrefmark{1}, Ge Gao\textsuperscript{\dag}\IEEEauthorrefmark{1}, Heming Sun\textsuperscript{\S}, Fan Zhang\textsuperscript{\dag} and David Bull\textsuperscript{\dag}}
\IEEEauthorblockA{  \textsuperscript{\dag}\textit{Visual Information Laboratory, University of Bristol, Bristol, UK, BS1 5DD} \\
  \{ha21615, ge1.gao, fan.zhang, dave.bull\}@bristol.ac.uk  \\
  \textsuperscript{\S}\textit{Yokohama National University, Japan} \\
  sun-heming-vg@ynu.ac.jp}
}

\maketitle

\begin{abstract}
In recent years, end-to-end learnt video codecs have demonstrated their potential to compete with conventional coding algorithms in term of compression efficiency. However, most learning-based video compression models are associated with high computational complexity and latency, in particular at the decoder side, which limits their deployment in practical applications. In this paper, we present a novel model-agnostic pruning scheme based on  gradient decay and adaptive layer-wise distillation. Gradient decay enhances parameter exploration during sparsification whilst preventing runaway sparsity and is superior to the standard Straight-Through Estimation. The adaptive layer-wise distillation regulates the sparse training in various stages based on the distortion of intermediate features. This stage-wise design efficiently updates parameters with minimal computational overhead. The proposed approach has been applied to three popular end-to-end learnt video codecs, FVC, DCVC, and DCVC-HEM. Results confirm that our method yields up to 65\% reduction in MACs and 2$\times$ speed-up with less than 0.3dB drop in BD-PSNR. Supporting code and supplementary material can be downloaded from: \url{https://jasminepp.github.io/lightweightdvc/}.
\end{abstract}


\section{Introduction}

With advances in mobile devices and video streaming services alongside the proliferation of user generated content, we have seen a significant increase in video consumption over the internet, further increasing tensions between data demand and limited transmission bandwidth \cite{bull2021intelligent}. Video compression algorithms are thus key in addressing this issue, with the most recent MPEG standard, H.266/Versatile Video Coding (VVC) \cite{VVC}, achieving a 30-40\% compression efficiency improvement \cite{bross2021overview} over its predecessor, H.265/High Efficiency Video Coding (HEVC) \cite{HEVC}. These improvements have largely been obtained through the development of new sophisticated coding tools within the conventional codec framework.

\begin{figure}[htbp]
    \centering
    \includegraphics[width=0.495\textwidth]{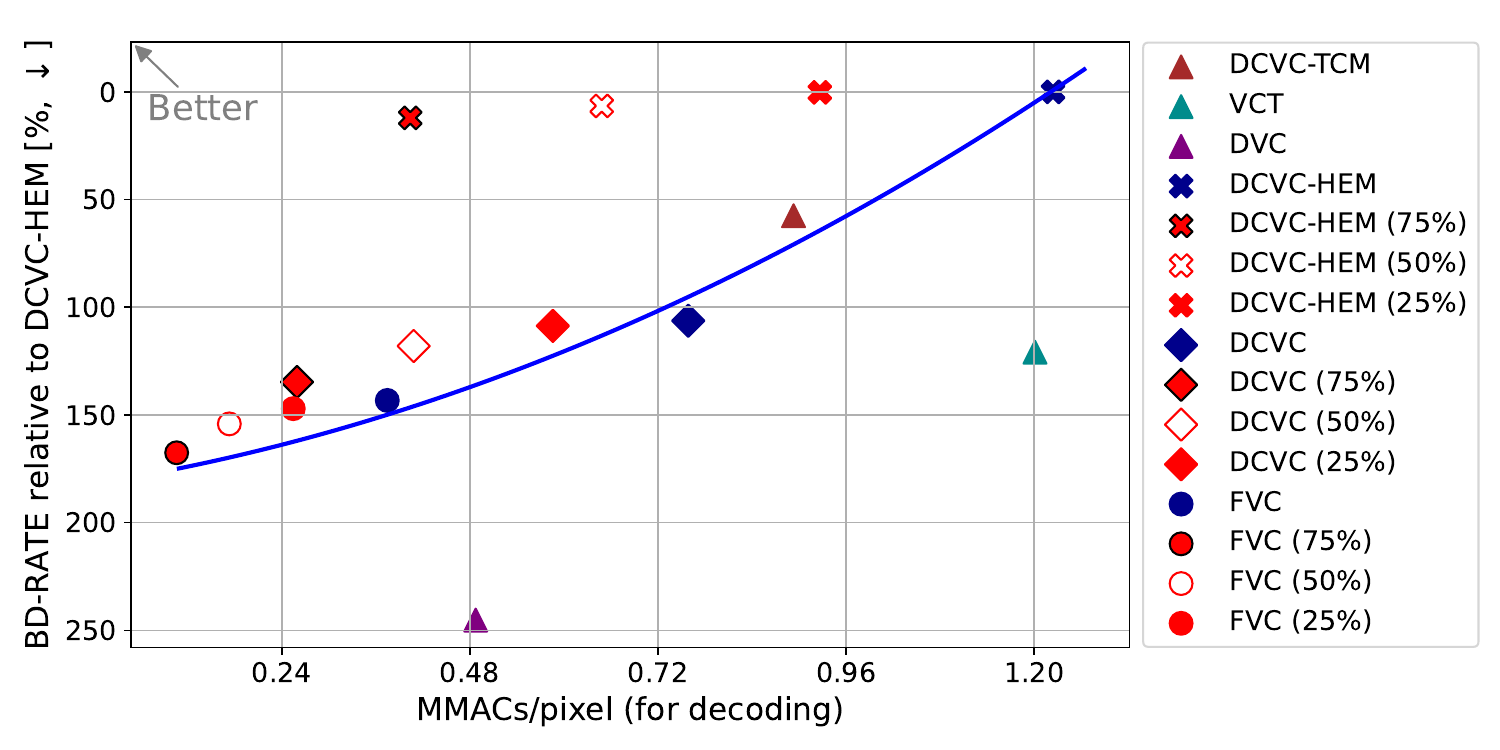}
    \caption{Comparing the performance-complexity trade-offs on UVG for different learnt video codecs. Models pruned using our method are positioned on the upper-left side of the frontier estimated from the original models.}
 \label{fig:plot}
   \vspace{-10pt}
\end{figure}

Recently, inspired by breakthroughs in deep learning, many enhancements to conventional coding tools have been proposed. These include in-loop filters \cite{ma2020mfrnet} post processing \cite{feng2022vistra3}, super-resolution \cite{ma2019perceptually}, etc. Alongside these, a new coding framework has emerged based on end-to-end optimisation \cite{lu2019dvc} with  the latest contributions \cite{hu2021fvc,li2021deep, gao2021neural, li2022hybrid,li2023neural, kwan2023hinerv} demonstrating significant potential to compete with the best standardised codecs. However, these end-to-end compression models are often associated with high encoding and decoding complexities, which limit their practical deployment in resource-constrained scenarios. The high computational and memory requirements with these new models also lead to increased carbon footprint and result in negative social and environmental impacts.

\begin{figure*}[htbp]
    \centering
    \includegraphics[width=\linewidth]{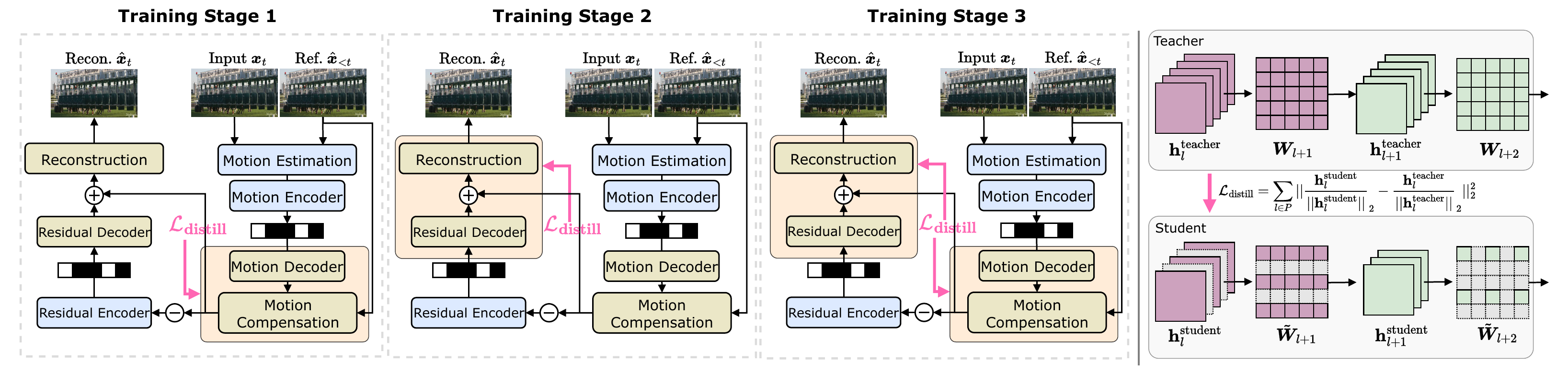}
    \caption{\textbf{Left}: Illustration of the iterative pruning process with layer-wise distillation. Only the modules on the decoder side, denoted in \textcolor{beige-color}{\textbf{beige}} colour, are selected for structured pruning. In each stage, the network is pruned alongside the distillation of a subset of modules. \textbf{Right}: Structured pruning with normalised feature distillation.}
    \label{fig:network}
\end{figure*}

To address this issue, pruning \cite{evc, luo2022memory, yin2022exploring} and quantisation \cite{sun2022q} techniques have been adopted for complexity reduction. Here, \textit{structured pruning}, which removes entire filters of neural nets, has been frequently applied, since it can achieve inference acceleration without requiring specific hardware or library support. However, this coarse-grained reduction of model parameters can also result in a significant performance loss, so efforts have been devoted to refining structured pruning techniques and better understanding the performance-complexity trade-offs in the context of learnt image and video compression. Advances have been made  in reducing runtime latency with minimal impact on complexity \cite{yin2022exploring, luo2022memory, kim2020efficient}. Notably, \cite{evc} has advanced these efforts by augmenting the pruning process with knowledge distillation, achieving a further reduction in performance degradation. However, the rate-distortion-complexity optimality of structured pruning for neural video codecs \cite{hu2023complexity, liu2022slimmable} remains underexplored. Video can be much more challenging because of the complexity arising from the need to exploit spatio-temporal redundancies.

In this paper, we propose a new complexity reduction approach for agnostically pruning end-to-end learnt video codecs. We first identify the issues in structural pruning caused by the commonly used straight-through approximator: (i) the forward-backward mismatching especially at higher sparsity rates and (ii) the instability problem when the surrogate gradient remains constant in each training stage. A simple but effective alternative, dubbed \textbf{gradient decay}, is therefore proposed, where we progressively \textit{decay} the proxy gradients of pruned weights as the training progresses, allowing flexible exploration in the early stages and stabilise convergence in later stages. Moreover, it is noted that the training of learnt video compression, unlike in learnt image compression, is typically split into various stages, each of which corresponds to the optimisation of one or multiple sub-modules within the codec. However, this training strategy is not appropriate when model pruning is involved, as our target is to constrain the global sparsity of the model rather than that of each individual sub-module. To this end, we introduce an \textbf{adaptive layer-wise distillation strategy} to regularise the pruned, student model, which leverages feature-level distillation by instead splitting the \textit{distillation} of sub-modules into stages (the pruning operation remains global). This implicitly supervises the global sparsification process with the knowledge of the stage-wise trained teacher. This proposed approach has been employed to reduce model complexity for three popular learnt video codecs, FVC \cite{hu2021fvc}, DCVC \cite{li2021deep}, and DCVC-HEM \cite{li2022hybrid}. We demonstrate its effectiveness, achieving up to 65\% reduction in MACs and 50\% reduction in latency with only less than 0.3dB loss in BD-PSNR when compared to their original counterparts.

The rest of the paper is organised as follows. Section \ref{sec:method} described the proposed the pruning workflow based on gradient decay and layer-wise distillation. It has been applied to three anchor codecs, and the results are then presented in Section \ref{sec:results} alongside analysis and discussion. Finally, Section \ref{sec:conclusion} provides a summary of the paper and outlines the future work.

\section{Method}
\label{sec:method}

Inspired by the research findings in recent studies \cite{evc, randomchannel, fantasticweights}, which indicate that different pruning criteria tend to yield similar results and that the focus should be instead on the optimisation method, we propose a simple but effective pruning scheme which deploys progressively decaying proxy gradients and adaptive layer-wise distillation to improve the training process. In this work, our complexity reduction focus is solely on the decoding part, which is essential to enable real-time playback. 

\subsection{Overview}
Fig. \ref{fig:network} illustrates how we iteratively prune the sub-modules in the decoder of a learnt video codec. The learning-based video compression pipeline consists of two branches: a motion coding branch and a residual coding branch. The input frame $\boldsymbol{x}_t$ is compressed by encoding both motion information (i.e., the motion vectors) and prediction residuals by performing motion estimation between $\boldsymbol{x}_t$ and its reconstructed reference(s), $\hat{\boldsymbol{x}}_{<t}$. The reconstructed frame is denoted by $\hat{\boldsymbol{x}}_t$. In each training stage in Fig. \ref{fig:network}, we prune the whole network but only distill (in a layer-wise manner) a subset of modules from their corresponding pre-trained, dense counterparts. 

We define the weight matrix of the $l$-th layer of the model as $\mathbf{W}^{(k)}_l\in \mathbb{R}^{d^{\text{out}}_l \times d^{\text{in}}_l}$, where $d^{\text{out}}_l$ and $d^{\text{in}}_l$ denote the layer's output and input channel dimensions, respectively. At each training iteration $k$, the importance score of each row (corresponding to a convolutional filter) $\mathbf{W}^{(k)}_{l,[j,:]}$, where $j=1, 2, ...,d^{\text{out}}_l$, is measured (detailed in Section \ref{sec:decay}) and the row is removed/pruned if its importance score is below the layer's corresponding threshold $T^{(k)}_l$. Here $T^{(k)}_l$ is a learnable parameter used to control the layer's structural sparsity. By doing this, all prunable sub-modules are optimised to reduce the overall rate-distortion-sparsity loss. 

\subsection{Gradient Decay}
\label{sec:decay}
It is noted that the hard thresholding operation is non-differentiable and therefore necessitates the use of surrogate gradients to enable efficient gradient-based sparsification. The most common choice for surrogate gradients is Straight-Through Estimation (STE) \cite{bengio2013estimating}, which defines the gradient with respect to the dense weight that was then assigned to the gradient of the thresholding operator's output. This allows temporarily pruned/dead weights to receive gradient updates and to be re-activated, thereby allowing higher exploration flexibility in early training stages when the pruning pattern is not fixed. However, STE with identity gradient copy could be problematic at later training stages when the pruning pattern is empirically observed to be stable. This means that the pruned weights should receive attenuated or even no gradient updates to avoid the noise introduced by forward-backward mismatching (which further exaggerates as sparsity increases). 

To this end, we propose a new gradient approximator, called \textbf{gradient decay}, which is initialised in the same manner as STE (i.e., identity copying), but is progressively decayed/annealed as training progresses. This is inspired by the weight decay proposed in \cite{evc}. Specifically, let ${N}^{(k)}_{l, j} = ||\mathbf{W}^{(k)}_{l,[j,:]}||_\text{2}$ be the L2-norm importance score of the $j$-th weight row of layer $l$, the forward and backward of the weight row after pruning, denoted by $\tilde{\mathbf{W}}^{(k)}_{l,[j,:]}$, is expressed as: 
\begin{equation}\label{eq:1}
\begin{aligned}
    &\text{Forward: } \tilde{\mathbf{W}}^{(k)}_{l,[j,:]} = \left\{
        \begin{array}{lc}
          \mathbf{W}^{(k)}_{l,[j,:]} & \text{if } N^{(k)}_{l, j} - T^{(k)} > 0 \\
          \textbf{0} & \text{otherwise}
        \end{array} \right. \\
    &\text{Backward: } \nabla{\tilde{\mathbf{W}}^{(k)}_{l,[j,:]}} = \left\{ 
        \begin{array}{lc} 
          \textbf{1} & \text{if } N^{(k)}_{l, j} - T^{(k)} > 0 \\
          \boldsymbol{\beta} & \text{otherwise,}
        \end{array}\right.
\end{aligned}
\end{equation}
where $\boldsymbol{\beta} \in [0, 1]$ is the decay parameter used to regulate the impact of pruned weights. We use the sigmoid scheduler to configure the value of $\boldsymbol{\beta}$ as a function of the current training iteration $k$ and the total training steps $K$:
\begin{equation}\label{eq:decay}
    \boldsymbol{\beta} = 1 - \text{sigmoid}(L_0 + (L_1 - L_0) \cdot k/K),
\end{equation}
in which $L_1$ and $L_0$ are pre-defined constants, which are used to control the decay rate. The global sparsity $s^{(k)}$ at the $k$-th training iteration is estimated as:
\begin{equation}
    s^{(k)} = 1 - \frac{1}{L \times d^{\text{out}}_l} \sum^L_{l=1} \sum^{d^{\text{out}}_l}_{j=1} \text{sigmoid} (\tau (N^{(k)}_{l, j}- T^{(k)}_l)),    
\end{equation}
where $\text{sigmoid}(\cdot)$ serves as a smooth approximation to the actual sparsity, and $\tau$ is a constant that controls the ``steepness'' of the sigmoid curve. 

\subsection{Iterative Pruning with Adaptive Distillation}

The overall rate-distortion-sparsity loss for the iterative pruning process described above is:
\begin{equation}\label{eq:rds}
    \mathcal{L}_{\text{RDS}} = \underbrace{R_m + R_r}_{\text{rate}} + \lambda_1 \underbrace{D(\boldsymbol{x}_t, \hat{\boldsymbol{x}}_t)}_{\text{distortion}} + \lambda_2 \underbrace{||s^{(k)} - s^{(k)}_{\text{tar}}||_1}_{\text{sparsity}},
\end{equation}
where $R_m$ and $R_r$ denote the bit-cost of motion features and residual features respectively after entropy coding. $ \lambda_1$ and $ \lambda_2$ are constant coefficients which trade off the balance between different loss terms. $D$ is the distortion term (between $\boldsymbol{x}_t$ and $\hat{\boldsymbol{x}}_t$), and $s^{(k)}_\text{tar}$ is the target sparsity based on the cubic sparsity schedule \cite{lazarevich2021post}:
\begin{equation}\label{eq:schedule}
    s^{(k)}_\text{tar} = 
        \begin{cases} 
          s_{\text{tar}} + (1 - s_{\text{tar}})(1 - \frac{k}{0.7 \cdot K})^3 
          & k < 0.7 \cdot K  \\
          s_{\text{tar}} & k \geq 0.7 \cdot K,
        \end{cases}  \\
\end{equation}    
where $s_{\text{tar}}$ is the global/final sparsity level. 
    
We empirically found that the sparse training process could also benefit from the progressive strategy \cite{li2021deep} that is commonly used when training neural video compression methods. For example, \cite{li2021deep}  first trains only the motion-related sub-modules with $R_m + D$, then the remaining modules with $D$, and finally the whole network with the full RD-loss. However, in our case, naively adopting this multi-stage training strategy is problematic, as the pruning of sub-modules in isolation at each training stage, is only locally optimal and won't align with the global sparsity constraint. Essentially, the inter-dependency of the modules requires a holistic training approach to achieve global sparsity by pruning all sub-modules together.

To address this issue, rather than \textit{explicitly} selecting sub-modules for individual training, a \textbf{layer-wise distillation strategy} is proposed where the sub-modules are \textit{implicitly} regularised through staged distillation with the pre-trained teacher model. This approach allows the sparse student model to inherit structured knowledge, ensuring that the global sparsity constraint is adhered without directly partitioning the training process. Further, this also reduces computation and memory overhead induced by full layer-wise distillation, making it particularly useful for neural video compression models that tend to be complicated and resource-demanding in the training process. The distillation loss at each is the sum of mean-squared loss between the normalised hidden states:
\begin{equation}\label{eq:distill}
    \mathcal{L}_{\text{distill}} = \sum_{l \in P} ||\frac{\textbf{h}_{l}^{\text{student}}}{||\textbf{h}_{l}^{\text{student}}||}_2 - \frac{\textbf{h}_{l}^{\text{teacher}}}{||\textbf{h}_{l}^{\text{teacher}}||}_2||^2_2,
\end{equation}
where $P$ denotes the indices of active layers at this stage, and $\textbf{h}_{l}^{\text{student}}$ and $\textbf{h}_{l}^{\text{teacher}}$ stand for the output of the student's and teacher's $l$-th layer, respectively. The overall loss at this stage is therefore expressed as:
\begin{equation}\label{eq:3}
    \mathcal{L}_{\text{all}} = \mathcal{L}_{\text{RDS}} + \lambda_3 \mathcal{L}_{\text{distill}}.
\end{equation}
Here $\lambda_3$ is also a hyperparameter controlling the trade off between $\mathcal{L}_{\text{RDS}}$ and $\mathcal{L}_{\text{distill}}$, which is progressively decayed with the same formular as Eq. (\ref{eq:decay}). This can effectively avoid over-compensation for distillation losses, otherwise the direction of parameter updates may diverge from that to $\mathcal{L}_{\text{RDS}}$ due to the increased capacity gap. 

\section{Experiments}
\label{sec:results}

\begin{figure*}[htbp]
    \centering
    \includegraphics[width=0.80\linewidth]{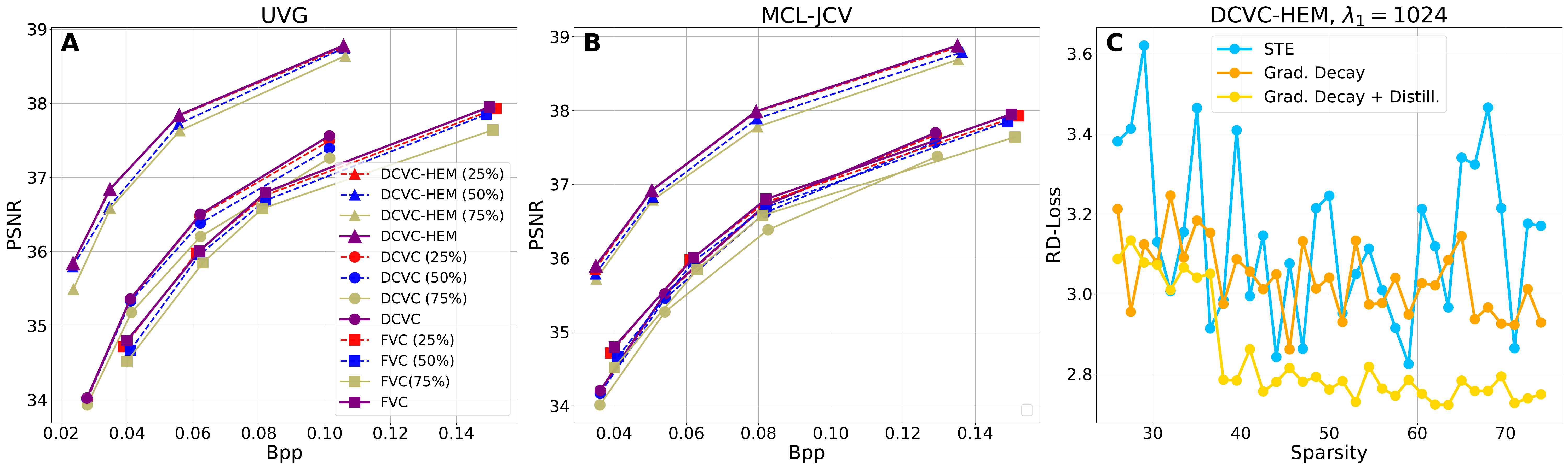}
    \caption{\textbf{A}-\textbf{B}: RD performance, measured by PSNR, of compact models against the original dense models on UVG and MCL-JCV, respectively; \textbf{C}: impact of gradient decay and layer-wise distillation on improving sparse training.}
    \label{fig:rd-plot}
\end{figure*}

We followed the common practice in training learning-based video codecs in this work, and employ the training part of the Vimeo-90K \cite{xue2019video} septuplet dataset. We adopted batch size of $4$ and a base training steps of $2\times 10^5$, corresponding to the dense model. We scale down the total training steps according to the target sparsity. The learning rate is initialised to $10^{-4}$ and progressively halved following a piecewise decaying schedule. $\lambda_1$ is set to 256, 512, 1024, and 2048, $\lambda_2$ is set to 20, and $\lambda_3$ is initialised to 1. MSE is used for the distortion loss $D$ in Eq. (\ref{eq:rds}).

To demonstrate the effectiveness of the proposed method, we select and prune three notable end-to-end learnt video codecs, FVC \cite{hu2021fvc}, DCVC \cite{li2021deep}, and DCVC-HEM \cite{li2022hybrid}, towards \textbf{three} target sparsity levels, including 25\%, 50\%, and 75\%, solely focusing on their decoders. Additionally, we follow \cite{hu2023complexity} to simplify FVC by removing the multi-frame feature fusion module and DCVC by removing the auto-regressive context model. We further compared the performance-complexity trade-off of our compact models to DCVC-TCM \cite{sheng2022temporal} and VCT \cite{mentzer2022vct} as shown in Fig. \ref{fig:plot}.

We employed the HD (1080p) test sequences from the UVG \cite{mercat2020uvg} and MCL-JCV \cite{wang2016mcl} databases to evaluate the compression performance of all the tested methods. To enable a fair comparison, we set the Group of Pictures size to 32 and adhered to the dense model's approach for encoding I frames. We use peak-to-noise ratio (PSNR) to measure video quality in this experiment. The average decoding latency (runtime) of HD P-frames are estimated using a NVIDIA 3090 GPU including \textit{arithmetic coding} (performed on an Intel Core i7-12700 CPU) but excluding disk I/O time.

As shown in Fig. \ref{fig:rd-plot}.(\textbf{A}-\textbf{B}) and TABLE \ref{tab:summary}, after pruning at different sparsity levels, the resulting compact models can still achieve competitive rate-distortion performance (even at the highest structural sparsity), but with a significant complexity reduction. For example, DCVC-HEM at 75\% sparsity is associated with a decoding latency reduction of 51\% but only at a 0.24dB drop in BD-PSNR on UVG and 0.29dB on MCL-JCV. It is noted that the decrease in decoding latency, while significant, is less pronounced compared to the more substantial reductions achieved in the number of parameters (i.e., model size) and MACs. This is largely attributed to the presence of I/O-intensive operations (e.g., warping) commonly seen in neural video codecs and the arithmetic coding process that is time-consuming, and the latency is expected to reduce further with the skip entropy coding techniques \cite{hu2023complexity}.

\begin{table}[htbp]
    \centering
    \caption{Average rate-distortion-complexity performance, evaluated on UVG and MCL-JCV dataset @1080P. All complexity figures are measured for decoders only. BD-Rate \& BD-PSNR are calculated against the corresponding dense model.}
    \label{tab:summary}
    \resizebox{1\linewidth}{!}{
    \begin{tabular}{l|c|c|c|c|c}
    \toprule
     \multicolumn{1}{l}{Metrics} & \multicolumn{1}{l}{Dataset} & \multicolumn{4}{c}{Sparsity Level} \\
    \cmidrule{3-6}
     \multicolumn{1}{l}{} & \multicolumn{1}{l}{} & \multicolumn{1}{c}{Original} & \multicolumn{1}{c}{25\%} & 
    \multicolumn{1}{c}{50\%} & \multicolumn{1}{c}{75\%} \\ 
    \midrule
     \multicolumn{6}{c}{\textbf{FVC}} \\
     \midrule
     \multirow{2}*{BD-PSNR/} & UVG & 0dB/0\% & -0.02dB/+0.80\% & -0.12dB/+4.02\% & -0.26dB/+10.10\% \\
                            \cmidrule{2-6}
                            BD-Rate & MCL & 0dB/0\% & -0.01dB/+0.23\% & -0.19dB/+6.22\% & -0.29dB/+11.55\% \\
     \cmidrule{1-6}
     Latency & / & 148ms & 134ms(\textcolor{blue}{$\downarrow11\%$}) & 92ms(\textcolor{blue}{$\downarrow39\%$}) & 73ms(\textcolor{blue}{$\downarrow51\%$})  \\
     Model Size & / & 7.36M & 5.59M(\textcolor{blue}{$\downarrow24\%$}) & 3.69M(\textcolor{blue}{$\downarrow50\%$}) & 2.28M(\textcolor{blue}{$\downarrow70\%$}) \\
     MACs& / & 0.68T & 0.53T(\textcolor{blue}{$\downarrow22\%$}) & 0.36T(\textcolor{blue}{$\downarrow47\%$}) & 0.22T(\textcolor{blue}{$\downarrow68\%$}) \\
     
     \midrule
     \multicolumn{6}{c}{\textbf{DCVC}} \\
     \midrule
     \multirow{2}*{BD-PSNR/} & UVG & 0dB/0\% & -0.02dB/+0.90\% & -0.08dB/+3.04\% & -0.23dB/+9.49\% \\
                            \cmidrule{2-6}
                            BD-Rate & MCL & 0dB/0\% & -0.01dB/+0.54\% & -0.10dB/+3.47\% & -0.29dB/+11.08\% \\
     \cmidrule{1-6}                             
     Latency & / & 296ms & 267ms(\textbf{\textcolor{blue}{$\downarrow11\%$}}) & 203ms(\textcolor{blue}{$\downarrow32\%$}) & 172ms(\textcolor{blue}{$\downarrow43\%$}) \\           
     Model Size & / & 3.78M & 3.03M(\textcolor{blue}{$\downarrow22\%$}) & 2.13M(\textcolor{blue}{$\downarrow44\%$}) & 1.22M(\textcolor{blue}{$\downarrow68\%$}) \\
     MACs & / & 1.58T & 1.22T(\textcolor{blue}{$\downarrow23\%$}) & 0.85T(\textcolor{blue}{$\downarrow46\%$}) & 0.54T(\textcolor{blue}{$\downarrow66\%$}) \\
     
     \midrule
     \multicolumn{6}{c}{\textbf{DCVC-HEM}} \\
     \midrule
     \multirow{2}*{BD-PSNR/} & UVG & 0dB/0\% & -0.01dB/+0.23\% & -0.11dB/+5.84\% & -0.24dB/+10.9\% \\
                            \cmidrule{2-6}
                            BD-Rate & MCL & 0dB/0\% & -0.01dB/+0.43\% & -0.11dB/+5.88\% & -0.29dB/+11.51\% \\
     \cmidrule{1-6}
     Latency & / & 519ms & 459ms(\textcolor{blue}{$\downarrow13\%$}) & 332ms(\textcolor{blue}{$\downarrow36\%$}) & 265ms(\textcolor{blue}{$\downarrow51\%$}) \\
     Model Size & / & 4.85M & 3.69M(\textcolor{blue}{$\downarrow24\%$}) & 2.72M(\textcolor{blue}{$\downarrow44\%$}) & 1.51M(\textcolor{blue}{$\downarrow69\%$}) \\
     MACs & / & 2.55T & 1.93T(\textcolor{blue}{$\downarrow24\%$}) & 1.35T(\textcolor{blue}{$\downarrow47\%$}) & 0.84T(\textcolor{blue}{$\downarrow67\%$}) \\
    \bottomrule
    \end{tabular}
    }
\end{table}

\begin{table}[ht!]
    \centering
    \caption{Ablation studies on BD-rate (\%, $\downarrow$) of DCVC on UVG.}
    \label{tab:ablation}
    \resizebox{0.75\linewidth}{!}{
    \begin{tabular}{r|ccc}
    \toprule
    Method & 25\% & 50\% & 75\% \\
    \midrule
    STE (v1) & 1.98& 5.41& 15.0\\ 
    GD (v2) & 0.98 & 4.57 & 12.2 \\ 
    STE and ALD (v3)  & 1.02 & 4.91 & 10.7 \\
    GD and full distillation (v4) & \cellcolor{lightgray}\textbf{0.89} & 3.06 & 9.55 \\
    \textbf{GD and ALD. (Ours}) & 0.90 & \cellcolor{lightgray}\textbf{3.04} & \cellcolor{lightgray}\textbf{9.49}  \\
    \bottomrule
    \end{tabular}
    }
    
\end{table}

A more intuitive illustration of the trade-off is shown in Fig. \ref{fig:plot}, where the compact models generated by our approach provide an excellent trade off between the coding efficiency and computational complexity (i.e. MACs/pixel) against benchmarked baselines on the UVG dataset. This is visualised by the positioning of all pruned models to the top left of the established rare-distortion-complexity boundary.

The ablative results in TABLE \ref{tab:ablation} show the effectiveness of two major contributions in this work, \textbf{gradient decay (GD)} and \textbf{adaptive layer-wise distillation (ALD)}. By directly comparing STE and GD for gradient approximation (w/o distillation), GD (v2) improves the DCVC model by 1.55\% (the average over three levels) over the standard STE (v1). This has also been confirmed when we involve the same distillation approach (ALD) together with different gradient approximator - v3 versus ours. Moreover, with our adaptive distillation method (ALD), the performance has been improved by  1.44\% (average) over v2 (GD only), and by 0.02\% over v4 (GD and full distillation). Here full distillation indicates that we apply knowledge distillation to all sub-modules regardless the stages \cite{lazarevich2021post}, which leads to much slower training speed. The excellent performance of these two contributions have also been manifested in Fig. \ref{fig:rd-plot}.(C), where training loss decreases more smoothly with gradient decay and layer-wise distillation.

 
\section{Conclusion}
\label{sec:conclusion}
This paper presents a generic workflow for complexity reduction in the context of learnt video compression, tackling the challenges of high complexity and decoding latency. Our method combines a novel gradient approximator with decay and an adaptive layer-wise distillation, enhancing adaptability and preventing runaway updates during sparsification. We demonstrate that this approach can reduce FLOPs and decoding time by 50\% with less than 0.3dB drop in BD-PSNR across three tested codecs. Future research should explore more granular structured pruning (e.g., N:M sparsity) and evaluate different layer-wise distillation metrics like KL-divergence.

\vspace{12pt}

\small
\bibliographystyle{IEEEtran}
\bibliography{Ref}

\end{document}